\documentclass[journal]{IEEEtran}
\usepackage[utf8]{inputenc}
\usepackage{float}
\usepackage{authblk}
\usepackage{chngcntr}
\usepackage{graphicx}
\usepackage{amssymb}
\usepackage{amsmath, amssymb}
\usepackage{placeins}
\usepackage{ragged2e}
\usepackage{tabularx}
\usepackage{array}
\usepackage[numbers,sort&compress]{natbib}
\DeclareMathOperator*{\E}{\mathbb{E}} 

\title{Quantifying Privacy in Nuclear Warhead Authentication Protocols}
\author[1]{Ruaridh R.\ Macdonald}
\author[1]{ R.\ Scott Kemp}
\affil[1]{Massachusetts Institute of Technology, Department of Nuclear Science and Engineering}
\date{\vspace{-5ex}}

\begin{document}

\maketitle

\begin{abstract}
International verification of nuclear warheads is a practical problem in which the protection of secret warhead information is of paramount importance. We propose a measure that would enable a weapon owner to evaluate the privacy of a proposed protocol in a technology-neutral fashion. We show the problem is reducible to `natural' and `corrective' learning. The natural learning can be computed without assumptions about the inspector, while the corrective learning accounts for the inspector's prior knowledge. The natural learning provides the warhead owner a useful lower bound on the information leaked by the proposed protocol. Using numerical examples, we demonstrate that the proposed measure correlates better with the accuracy of a maximum {\em a posteriori} probability estimate than alternative measures.
\end{abstract}

\section{Introduction}

The core technical challenge facing future nuclear warhead disarmament treaties is how to verify that warheads are being irreversibly dismantled without revealing confidential warhead-design information to an inspector. Nation states worry that revealing their nuclear-weapon design secrets may compromise their nuclear deterrents, for example, by allowing adversaries to fine-tune countermeasures. However, without rigorous disarmament verification the disarming state may cheat by dismantling ``hoax objects'' while retaining their authentic warheads in a secret reserve \cite{bunch2014supporting, Bowen2018, fuller2010verification, DOE_dismantlement_verif}.

Concern about information leakage has prevented the practical use of warhead authentication for more than fifty years. Protocols with significantly enhanced privacy have been more recently proposed \cite{glaser_zero-knowledge_2014,kemp_physical_2016,marleau_investigation_2017, gilbert_single-pixel_2017}, but there are no technology-neutral measures for comparing their relative merits. Protocols need to be evaluated on their ability to correctly accept authentic warheads (completeness), reject hoax objects (soundness), and not reveal secret information (privacy). In this paper we propose a measure of privacy based on how much an inspector learns when executing the protocol. Given a set of assumptions about which warhead design details are important to protect, and the prior knowledge of the inspector; our measure can be used to compare competing authentication concepts.

\section{Privacy measure for Warhead Verification}

We define nuclear warheads by a set of properties, denoted $\Theta$. These could include isotope distributions, detonation yields, or other properties. Due to manufacturing variation, each warhead can be slightly different so $\Theta$ is a random variable with alphabet $\vartheta$ and specific instances $\theta\in\vartheta$, distributed according to $p_{\text{true}}(\Theta)$. This true distribution is only known to the warhead owner. The inspector starts with a prior belief distribution, $p_0(\Theta)$, which is their best estimate of $p_{\text{true}}(\Theta)$. The inspector updates their beliefs using the measured data, $X$, according to Bayes' theorem, shown in Equation~\ref{eq:bayes_law}. The numbered subscripts on $p$ denote the number of completed measurements. 

{\footnotesize
\begin{equation}
 p_1(\Theta=\theta) = p_0\left(\Theta=\theta| X_1=x\right) = \frac{p_0(x|\theta)p_0(\theta)}{p_0(x)}
 \label{eq:bayes_law}
\end{equation}
}

\noindent The data, $X$, is also a random variable, with alphabet $\mathcal{X}$ and possible results $x\in\mathcal{X}$, distributed according to  $p_{\text{true}}(X)$. It is dependent on the properties of the specific warhead being measured as well as the stochastic variation of the measurement. The stochastic variation is described by the likelihood function $p_0(x|\theta)$, which only depends on the physics and procedure of the measurement process. We will assume that both the owner and inspector understand the protocol perfectly, so $p_0(x|\theta)$ is known exactly and for that reason:

{\footnotesize
\begin{equation}
 p_0(x|\theta) = p_1(x|\theta) = p_2(x|\theta) = ... = p_{\text{true}}(x|\theta)
 \label{eq:privacy_likelihood}
\end{equation}
}

To assess privacy, we need a means of quantifying how much a measurement improves the accuracy of the inspector's beliefs about $\Theta$. There are multiple means of measuring the change between $p_0(\Theta)$ and $p_1(\Theta)$, but different measures highlight different features of any change. 

The most common approach is to estimate learning by the reduction in the entropy of the inspector's beliefs after conditioning on the measured data, as originally developed by Shannon, Lindley, and others\cite{lindley1956measure, kullback1997information, soofi1990effects}, with the average reduction being described by the mutual information between the property of interest and observable data. However, the entropy-based approach suffers from three shortcomings which reduce its usefulness in this application. Mutual information estimates the average amount the inspector learns about $\Theta$ assuming their prior belief distribution was $p_{\text{true}}(\Theta)$. This will not (indeed, must not) be generally true in warhead authentication, otherwise there is no private information, which is an essential feature of the problem. Furthermore, a reduction in Shannon or differential entropy reflects reduction in uncertainty about a random variable, rather than necessarily a reduction in inaccuracy. If the inspector becomes convinced of an incorrect value for $\Theta$, the entropy of their belief distribution will still decrease, but the owner would not consider convergence to a wrong answer to be demonstrative of weak privacy. Finally, the fact that Shannon entropy (though not the mutual information) can be negative or undefined for non-discrete random variables makes comparisons more difficult.
    
\subsection{Quantifying changes in knowledge by changes in KL divergences}
We propose to evaluate the inspector's learning by calculating how much conditioning on the measured data reduces the Kullback-Leibler divergence between $p_{\text{true}}(\Theta)$ and the inspectors belief distribution. We quantify the (in)accuracy of the inspector beliefs about $\Theta$ as the KL divergence between their belief distribution for $\Theta$ and the true distribution for $\Theta$. Their initial inaccuracy is given by Equation~\ref{eq:initial_fault}. The smaller this KL divergence is, the more accurate the inspector's initial belief distribution. This initial value is independent of the authentication protocol being analyzed. As long as the same properties are being considered, this is a consistent starting point for comparing protocols. 

{\footnotesize
\begin{equation}
D_{KL}(p_{\text{true}}(\Theta)||p_0(\Theta)) = \sum_{\theta\in\vartheta} p_{\text{true}}(\theta)\log_2\left(\frac{p_{\text{true}}(\theta)}{p_0(\theta)}\right)
\label{eq:initial_fault}
\end{equation}
}

\noindent After a measurement $X$, the inaccuracy in the inspector's belief is:

{\footnotesize
\begin{align}
D_{KL}(p_{\text{true}}(\Theta)||p_1(\Theta)) &= D_{KL}(p_{\text{true}}(\Theta)||p_0(\Theta|x)) \\
&= \sum_{\theta\in\vartheta} p_{\text{true}}(\theta)\log_2\left(\frac{p_{\text{true}}(\theta)}{p_0(\theta|x)}\right)
\label{eq:final_fault}
\end{align}
}

\noindent The change in the accuracy of the inspector's beliefs in $\Theta$ is the difference between Equations \ref{eq:initial_fault} and \ref{eq:final_fault}. If $x$ was a useful measurement, the belief inaccuracy will be reduced, thus we subtract the posterior error from the prior error to calculate the learning, as shown in Equations~\ref{eq:learning_1} to \ref{eq:learning_final}.

{\footnotesize
\begin{align}
\text{Learning} &\triangleq \mathcal{K}(p_0(\Theta),p_{\text{true}}(\Theta),x) \\
&= D_{KL}(p_{\text{true}}(\Theta)||p_0(\Theta))  - D_{KL}(p_{\text{true}}(\Theta)||p_1(\Theta))\label{eq:learning_1} \\
&= \sum_{\theta\in\vartheta}p_{\text{true}}(\theta) \log_2\left(\frac{p_1(\theta)}{p_0(\theta)}\right) \\
&= \sum_{\theta\in\vartheta}p_{\text{true}}(\theta) \log_2\left(\frac{p_0(\theta|x)}{p_0(\theta)}\right)\label{eq:learning_final}
\end{align}
}

\noindent The learning measure in \ref{eq:learning_final} incorporates the inspector's initial knowledge $p_0(\Theta)$, the data available to them $x$, and the description of the true warhead properties $p_{\text{true}}(\Theta)$. The inspector's method of inference is implied in the posterior belief distribution $p_0(\Theta|x)$. The owner does not have access to all of these components; in most cases the owner will have to guess the inspector's prior information and inference method. We will see the impact of these approximations in the following subsections. Note that if $p_0(\Theta) = p_{\text{true}}(\Theta)$, then Equation~\ref{eq:learning_final} will reduce to the mutual information when averaged over $x\in\mathcal{X}$. This suggests a strong link between our method and the traditional entropy-based approach, but ours accounts for the inspector's faulty prior.

While the learning measure in Equation~\ref{eq:learning_final} shares some similarities with Jensen-Shannon divergence and comparative divergence, the authors could not find it in the literature. This is probably because $p_{\text{true}}(\theta)$ is not known \textit{a priori} in most inference or learning situations. 

\section{Expectation value of learning}

To expand on the features of our proposed measure, in this section we calculate how much the owner should expect the inspector to learn about a class of warheads, based on one or more measurements.

\subsection{Measurement of one warhead}
To begin with, we assume the inspector takes one measurement of a single warhead. As above, the class of warheads has properties $\Theta$, distributed according to $p_{\text{true}}(\Theta)$. The specific warhead which the inspector measures has $\Theta = \theta^*$:

{\footnotesize
\begin{equation}
p_{\text{measured warhead}}(\Theta) = p_{\text{true}}(\Theta|\theta^*) = \begin{cases}
                                                                1, &\text{if } \Theta = \theta^* \\
                                                                0, &\text{otherwise }
                                                               \end{cases}
\end{equation}
}

\noindent Therefore, after measuring $x$, the inspector will have learned:

{\footnotesize
\begin{align}
 \mathcal{K}(p_0(\Theta), p_{\text{true}}(\Theta|\theta^*),x|\theta^*) &= \sum_{\theta\in\vartheta}p_{\text{true}}(\theta) \log_2\left(\frac{p_0(\theta|x)}{p_0(\theta)}\right)  \\
 &= 1\times\log_2\left(\frac{p_0(\theta^*|x)}{p_0(\theta^*)}\right)
\end{align}
}

\noindent We can find the expected result by averaging over all the possible measurement results:

{\footnotesize
\begin{align}
\mathcal{K}(p_0(\Theta), p_{\text{true}}(\Theta|\theta^*),X|\theta^*) &= \E_{x\in\mathcal{X}}\left\{ \mathcal{K}(p_0(\Theta),p_{\text{true}}(\Theta|\theta^*),x|\theta^*) \right\}\\
 &= \sum_{x\in\mathcal{X}} p_{\text{true}}(x|\theta^*)\log_2\left(\frac{p_0(\theta^*|x)}{p_0(\theta^*)}\right)
\end{align}
}

\noindent Using Bayes' law (Equation~\ref{eq:bayes_law}), we can trade the argument of the logarithm:

{\footnotesize
\begin{equation}
 \frac{p_0(\theta^*|x)}{p_0(\theta^*)} = \frac{p_0(x|\theta^*)}{p_0(x)}
\end{equation}
}

\noindent Remembering also that the likelihood function for $x|\theta^*$ is the same for $p_0$ and $p_{\text{true}}$ (Equation~\ref{eq:privacy_likelihood}), we can also exchange $p_{\text{true}}(x|\theta^*)$ for $p_0(x|\theta^*)$, or vice versa.

{\footnotesize
\begin{align}
\mathcal{K}(p_0(\Theta), p_{\text{true}}(\Theta|\theta^*),X|\theta^*) &= \sum_{x\in\mathcal{X}} p_0(x|\theta^*)\log_2\left(\frac{p_0(x|\theta^*)}{p_0(x)}\right) \\
 &= D_{KL}\left(p_0(x|\theta^*)||p_0(x)\right) \\
 &= D_{KL}\left(p_{\text{true}}(x|\theta^*)||p_0(x)\right)\label{eq:single_obj_learn}
\end{align}
}

\noindent The expected learning is a KL divergence, so it is non-negative. This tells us that, on average, the inspector's belief will be improved or stay the same after measuring the warhead. Individual samples of $x$ may be misleading and produce a negative learning measure, but the owner should expect that, on average, the inspector will learn something about $\Theta$ if it is related to $x$.

\subsection{Measurements over of a class of warheads}
We can average the result in Equation~\ref{eq:single_obj_learn} over the class of warheads to see how much the owner should expect the inspector to learn when measuring several different warheads.

{\footnotesize
\begin{align}
\mathcal{K}(p_0(\Theta), & p_{\text{true}}(\Theta),X) = \E_{\theta^*\in\Theta}\left\{ \mathcal{K}(p_0(\Theta),p_{\text{true}}(\Theta|\theta^*),X|\theta^*) \right\} \\
 &= \sum_{\theta^*\in\Theta} p_{\text{true}}(\theta^*) \sum_{x\in\mathcal{X}} p_{\text{true}}(x|\theta^*)\log_2\left(\frac{p_{\text{true}}(x|\theta^*)}{p_0(x)}\right) \\
 &= \sum_{\theta^*\in\Theta}\sum_{x\in\mathcal{X}} p_{\text{true}}(x,\theta^*)\log_2\left(\frac{p_{\text{true}}(x|\theta^*)}{p_0(x)}\right) \label{eq:learning_concise}
\end{align}
}

\noindent Using the definition of conditional and joint probabilities, we can rearrange the arguments of the logarithm again. At this stage, because we are averaging over $\Theta$, we can revert to using $\theta$ rather than $\theta^*$ to describe the warhead properties.

{\footnotesize
\begin{equation}
\mathcal{K}(p_0(\Theta), p_{\text{true}}(\Theta),X) 
    = \sum_{\theta\in\vartheta}\sum_{x\in\mathcal{X}} p_{\text{true}}(x,\theta) 
    \log_2\left(\frac{p_{\text{true}}(x,\theta)}{p_0(x)p_{\text{true}}(\theta)} \right)
\label{eq:learning_measure_intermediate}
\end{equation}
}

\noindent Note that Equation~\ref{eq:learning_measure_intermediate} is very similar to the expression for mutual information. Because the mutual information is purely a function of the warhead properties and measurement process, the owner can calculate it without having to make assumptions about the inspector. It would therefore be useful to describe the inspector's learning with respect to it.

{\footnotesize
\begin{align}
\begin{split}
    &\mathcal{K}(p_0(\Theta),p_{\text{true}}(\Theta),X) \\
    &= \sum_{\theta\in\vartheta}\sum_{x\in\mathcal{X}} p_{\text{true}}(x,\theta) 
    \log_2\left(\frac{p_{\text{true}}(x,\theta)}{p_0(x)p_{\text{true}}(\theta)}
    \frac{p_{\text{true}}(x)}{p_{\text{true}}(x)} \right)
\end{split} \\
\begin{split}
    &=\sum_{\theta\in\vartheta}\sum_{x\in\mathcal{X}} p_{\text{true}}(x,\theta) 
    \left( \log_2\left(\frac{p_{\text{true}}(x,\theta)}{p_{\text{true}}(x)p_{\text{true}}(\theta)}\right) \right. \\
    &\quad~~~~~~~~~~~~~~~~~~~~~~~~~~~~~~~~~~~~~~~~ \left.+ \log_2\left(\frac{p_{\text{true}}(x)}{p_0(x)}\right) \right)
\end{split}
\\
\begin{split}
    &=\sum_{\theta\in\vartheta}\sum_{x\in\mathcal{X}} p_{\text{true}}(x,\theta) 
    \log_2\left(\frac{p_{\text{true}}(x,\theta)}{p_{\text{true}}(x)p_{\text{true}}(\theta)}\right) \\
    &\quad~~~~~~~~~~~~~~~~~~~~~~~~~~+ \sum_{x\in\mathcal{X}}p_{\text{true}}(x)\log_2\left(\frac{p_{\text{true}}(x)}{p_0(x)}\right)
\end{split}
 \\
&= \text{I}(p_{\text{true}}(\Theta);p_{\text{true}}(X)) 
+ \sum_{x\in\mathcal{X}}p_{\text{true}}(x)\log_2\left(\frac{p_{\text{true}}(x)}{p_0(x)}\right)
 \\
&=\underbrace{\text{I}(p_{\text{true}}(\Theta);p_{\text{true}}(X))}_\text{\footnotesize natural learning} + \underbrace{D_{KL}(p_{\text{true}}(X)||p_0(X))}_\text{\footnotesize corrective learning} \label{eq:learning_measure}
\end{align}
}

\noindent Equation~\ref{eq:learning_measure} is the most important result. It can be used as the privacy measure for a parameter $\Theta$. There are two terms, a `natural learning' component, which depends on the true correlation between $\Theta$ and $X$; and a `corrective learning' component, which depends on how accurate the inspector's prior belief distribution is. The natural learning is described by the mutual information between $\Theta$ and $X$ under their true distributions. The owner can calculate it accurately without making assumptions about the inspector, other than Equation~\ref{eq:privacy_likelihood}. The corrective learning is the KL divergence between the true distribution of the output data, and the inspector's prior estimate of the same. It is highly dependent on the owner's estimate of the inspector's prior. 

The inspector's total learning will be greater than the natural learning because the corrective learning is a KL divergence and hence non-negative. Starting from a more inaccurate prior accelerates the rate of learning, because there is a greater difference between the measured data and what the inspector expected. The natural learning component is a useful lower bound on the average amount learned by the inspector, but no general upper bound exists. KL divergences are not upper-bounded in general, and can be infinite in some cases. However, maximizing the corrective learning will not leave the inspector with the most knowledge possible, i.e. minimize Equation~\ref{eq:final_fault}, in general. This is expanded on in section~\ref{sec:maxcorrective}. 

If the the inspector's prior is equal to the true distribution, the corrective learning will be zero and the total learning equal to the natural learning. This returns us to the typical entropy-based approach, again showing the relationship between it and our proposed measure. Inferring hyper parameters is discussed in more detail in section~\ref{sec:hyperparameters}.

\subsection{Effect of maximizing corrective learning}\label{sec:maxcorrective}

Choosing a very inaccurate prior belief distribution for the inspector will increase their average corrective learning. Here we show that choosing a prior to maximize the average corrective learning, $D_{KL}(p_{\text{true}}(X)||p_0(X))$, is not guaranteed to maximize the accuracy of the inspector's final belief distribution, $D_{KL}(p_{\text{true}}(\Theta)||p_0(\Theta|X))$. 

{\footnotesize
\begin{align}
D_{KL}(p_{\text{true}}(\Theta)||p_0(\Theta|X)) &= \E_{x\in\mathcal{X}}\left\{ D_{KL}(p_{\text{true}}(\Theta)||p_0(\Theta|x)) \right\}  \\ 
&= \sum_{x\in\mathcal{X}}p_{\text{true}}(x|\theta)\sum_{\theta\in\vartheta} p_{\text{true}}(\theta)\text{log}_2\left(\frac{p_{\text{true}}(\theta)}{p_0(\theta|x)}\right) \\
 &= \sum_{\theta\in\vartheta}\sum_{x\in\mathcal{X}} p_{\text{true}}(x,\theta)\text{log}_2\left(\frac{p_{\text{true}}(\theta)}{p_0(\theta|x)}\right)
\end{align}
}

\noindent Using Bayes law, we rewrite this equation in terms of the corrective learning. Reassuringly, we are able to rederive the learning measure from this alternate starting point.

{\footnotesize
\begin{align}
\begin{split}
    &\E_{x\in\mathcal{X}}\left\{ D_{KL}(p_{\text{true}}(\Theta)||p_0(\Theta|x)) \right\} \\
    &= \sum_{x\in\mathcal{X}, \theta\in\vartheta} p_{\text{true}}(x,\theta)\text{log}_2\left(\frac{p_{\text{true}}(\theta)}{p_0(\theta|x)}\right)
\end{split} \\
 &= \sum_{x\in\mathcal{X}, \theta\in\vartheta} p_{\text{true}}(x,\theta) \text{log}_2\left(p_{\text{true}}(\theta)\frac{p_0(x)}{p_0(x|\theta)p_0(\theta)}\right)\\
 &= \sum_{x\in\mathcal{X}, \theta\in\vartheta} p_{\text{true}}(x,\theta) \text{log}_2\left(\frac{p_{\text{true}}(\theta)p_0(x)}{p_0(x|\theta)p_0(\theta)}\frac{p_{\text{true}}(x)}{p_{\text{true}}(x)}\frac{p_{\text{true}}(\theta)}{p_{\text{true}}(\theta)}\right)\\
 \begin{split}
    &= \sum_{x\in\mathcal{X}, \theta\in\vartheta} p_{\text{true}}(x,\theta) \left( 
    \text{log}_2\left(\frac{p_{\text{true}}(\theta)}{p_0(\theta)}\right) \right.\\
    &\quad ~~~ \left.+ \text{log}_2\left(\frac{p_{\text{true}}(\theta)p_{\text{true}}(x)}{p_{\text{true}}(x|\theta)p_{\text{true}}(\theta)}\right)
    + \text{log}_2\left(\frac{p_0(x)}{p_{\text{true}}(x)}\right) \right) 
 \end{split}\\
 \begin{split}
     &= D_{KL}(p_{\text{true}}(\Theta)||p_0(\Theta)) - \text{I}(p_{\text{true}}(\Theta);p_{\text{true}}(X)) \\
     &\quad ~~~~~~~~~~~~~~~~~~~~~~~~~~~~ - D_{KL}(p_{\text{true}}(X)||p_0(X))\label{eq:worst_prior_check}
 \end{split} \\
&= D_{KL}(p_{\text{true}}(\Theta)||p_0(\Theta)) - \mathcal{K}(p_0(\Theta),p_{\text{true}}(\Theta),X) 
\end{align}
}

\noindent Looking at Equation \ref{eq:worst_prior_check}, we see that the accuracy of the inspector's final belief distribution depends on three factors: the accuracy of their prior belief in $\Theta$, the mutual information between $\Theta$ and the data, and the corrective learning. If we choose $p_0(\Theta)$ to maximize the corrective learning, we would expect the accuracy of the prior in $\Theta$ to decrease as well, and the first term of Equation \ref{eq:worst_prior_check} to increase, because the two are related by the likelihood function $p(x|\theta)$. It is simple to demonstrate by example that the first term can grow faster than the corrective learning term, showing that maximizing the corrective learning is not guaranteed to minimize the error in the inspector's final belief distribution.

\subsection{Multiple measurements of a warhead}

If the inspector is allowed to make multiple measurements of each test object then they will learn more about the warhead properties. The second result, $X_2$, is conditionally dependent on the first if $X_1$ provides any information about $\Theta$.  The learning measure for the second measurement is given by Equation~\ref{eq:secondMeasure}.

{\footnotesize
\begin{align}
\begin{split}
    \mathcal{K}( & p_0(\Theta|X_1),p_{\text{true}}(\Theta|X_1),X_2|X_1) \\
    &= \text{I}(p_{\text{true}}(\Theta|X_1);p_{\text{true}}(X_2|X_1)) + D_{KL}(p_{\text{true}}(X_2|X_1)||p_0(X_2|X_1))
\end{split}\\
    &= \sum_{\vartheta,\mathcal{X}_1,\mathcal{X}_2}p_{\text{true}}(x_1,x_2,\theta)\log_2\left(\frac{p_{\text{true}}(x_2|x_1,\theta)}{p_0(x_2|x_1)}\right)
    \label{eq:secondMeasure}
\end{align}
}

\noindent The average total learning over both measurements is the sum of Equations \ref{eq:learning_measure} and \ref{eq:secondMeasure}. Unless the measurement results are entirely independent Equation \ref{eq:secondMeasure} will be less than Equation \ref{eq:learning_measure}, and the inspector will face diminishing marginal returns with each successive measurement.

\subsection{Combining learning measures into utility functions}

Nuclear warheads are complicated objects and are best described by multiple properties. The owner can treat all of these together as one variable $\Theta$ and calculate a single value for the learning measure, in which case the privacy of the protocol is given by the learning measure: $\mathcal{K}(p_0(\Theta),p_{\text{true}}(\Theta),X)$. However, some properties may be more important for the owner to protect than others, in which case the warhead can be described by a set of properties $\mathcal{T}_N = \{\Theta_1,\Theta_2,...,\Theta_N\}$. The owner calculates the privacy measure for each property, and combines them in a privacy utility function. 

The owner can use any privacy utility function they wish, as long as they use the same one for all protocols. The utility function will embody a combination of political and technical concerns about which properties of a warhead are most important to protect, including emergent properties such as explosive yield. The owner is unlikely to share their utility function with the inspector, as it reflects sensitive information. In the simplest case, the privacy utility function could be a weighted sum of the individual privacy measures using fixed weights $\mathcal{W}_N = \{w_1,w_2,...,w_N\}$, as given in Equation~\ref{eq:utility_function}. A more sophisticated function would use weights which depend on the privacy measure values. 

{\footnotesize
\begin{equation}
    K(\mathcal{T}_N,\mathcal{W}_N,X) = \sum_{i=1:N}w_i\mathcal{K}(p_0(\Theta_i),p_{\text{true}}(\Theta_i),X)\label{eq:utility_function}
\end{equation}
}

\subsection{Numerical example}\label{sec:num_example}

In this section we provide a simple numerical example to demonstrate that our approach is a more accurate measure of inspector knowledge than Shannon entropy. The expression for the change in Shannon entropy is given in Equation~\ref{eq:entropy_change}.

{\footnotesize
\begin{equation}
    \E\{H(p_0(\Theta)) - H(p_0(\Theta|x))\} = \sum_{x\in\mathcal{X}}\sum_{\theta\in\vartheta}
    p_{\text{true}}(x)p_0(\theta|x)\log_2\left( \frac{p_0(\theta|x)}{p_0(\theta)} \right)
    \label{eq:entropy_change}
\end{equation}
}

For this example, we define warheads using one property, $\Theta$, distributed according to a normal distribution with mean $M$ and variance $S^2$. The inspector can take measurements of the test objects, producing data $X$, which we assume is distributed according to a Poisson distribution reflecting the statistics of frequently used radiation-based tools. Starting from a prior belief distribution $p_0(\Theta)$, the inspector uses a maximum \textit{a posteriori} probability (MAP) estimator for $\Theta$, which we denote $\Tilde{\Theta}$. $p_0(\Theta)$ is a discrete uniform distribution between two limits.

{\footnotesize
\begin{equation}
    p_{\text{true}}(\Theta=\theta) = \mathcal{N}[\theta,\mu=M,\sigma=S^2 ]
\end{equation}
\begin{equation}
    p_{\text{true}}(X=x|\theta) = \mathcal{P}oisson[x,\lambda=\theta ]
\end{equation}
}

\begin{figure}
    \centering
    \noindent\makebox[0.45\textwidth]{
    \includegraphics[width=0.45\textwidth,angle=-0,origin=c]{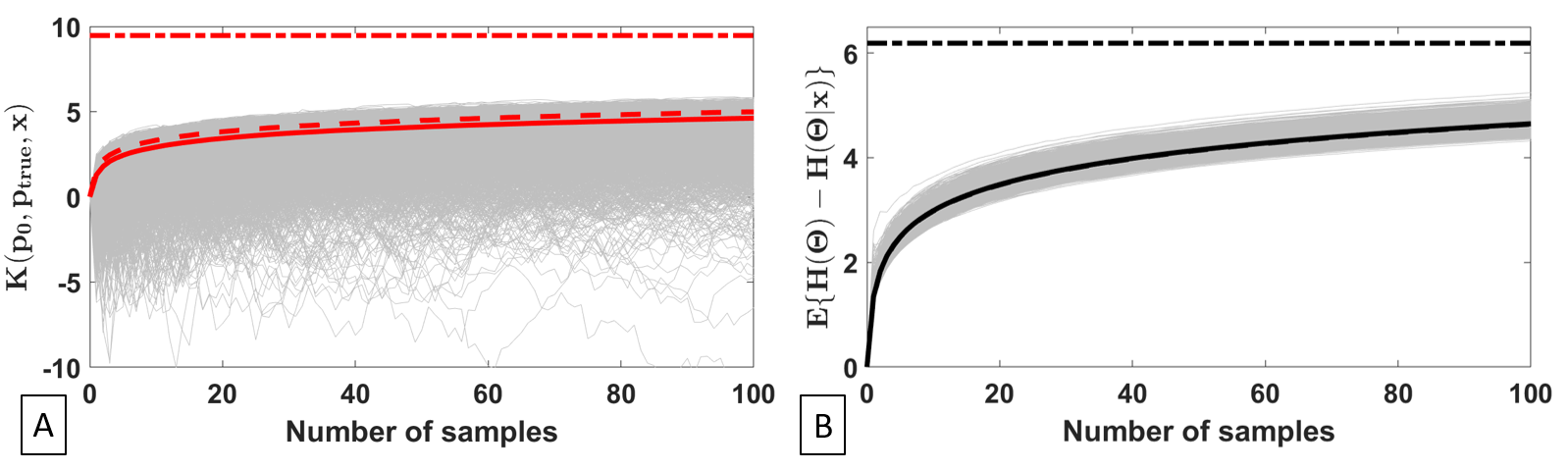}
    }
    \caption{Comparison of the measure proposed in this paper [A] and Shannon entropy [B] versus the number of measurements made. The simulation was run for 10,000 cases, with $M=40,S=7$. The learning measures for each individual case are shown in grey, the mean result by the solid curve, and the median result by the dashed curve. The median and mean result are almost identical in [B]. The KL divergence and Shannon entropy of the prior distribution are given by the dot-dashed lines.}
    \label{fig:numExample_1}
\end{figure}

\begin{figure}
    \centering
    \noindent\makebox[0.45\textwidth]{
    \includegraphics[width=0.45\textwidth,angle=-0,origin=c]{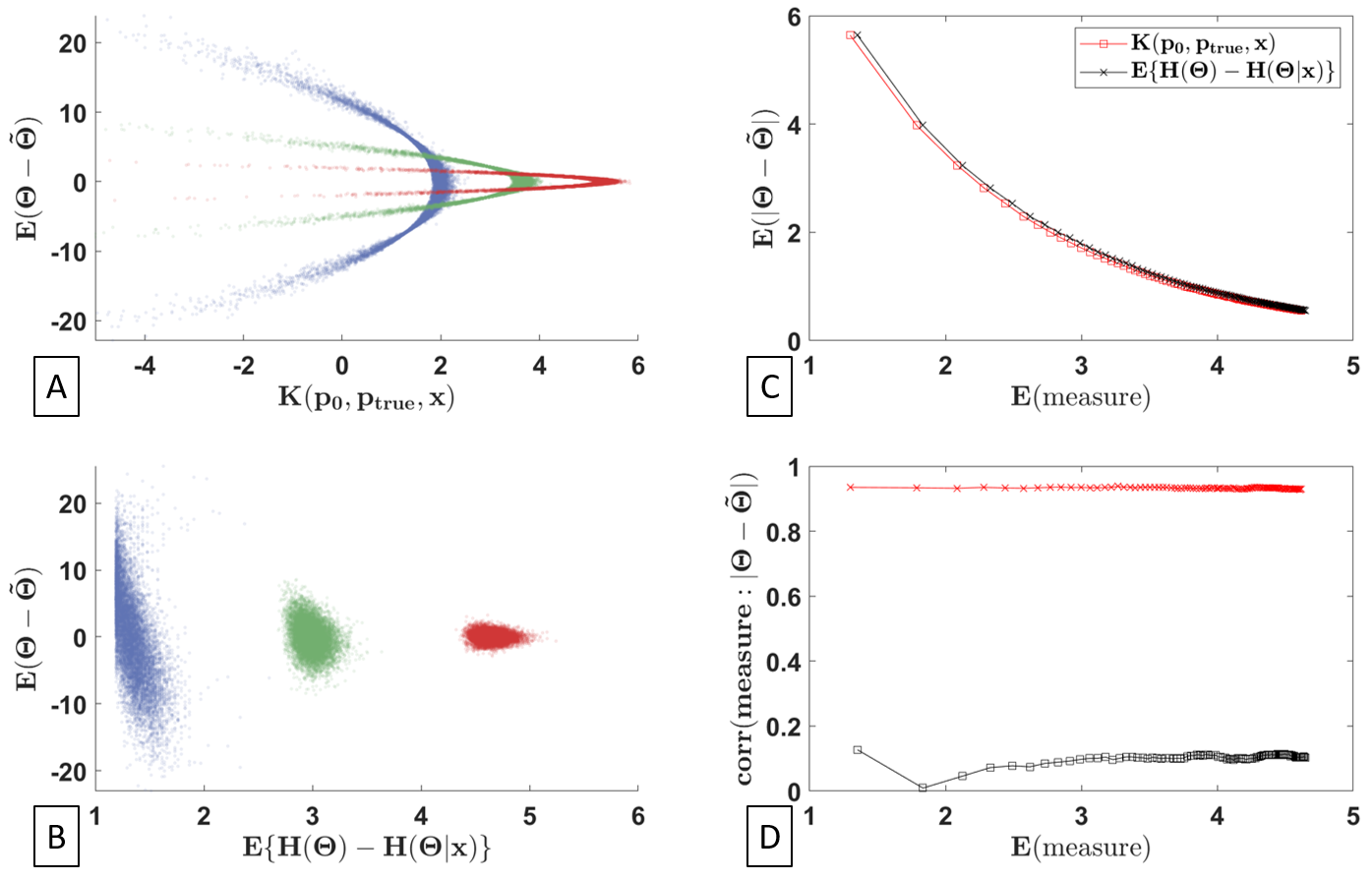}
    }
    \caption{Changes in the values of the two learning measures versus the MAP estimator error for $\Theta$. [A] and [B] show the result for each of the 10,000 cases, after 1, 10, and 100 measurements in blue, green, red respectively. [C] depicts the mean absolute MAP error versus both learning measures. [D] gives the linear correlation between the absolute MAP error and the two measures after each measurement.}
    \label{fig:numExample_2}
\end{figure}

\begin{figure}
    \centering
    \noindent\makebox[0.45\textwidth]{
    \includegraphics[width=0.45\textwidth,angle=-0,origin=c]{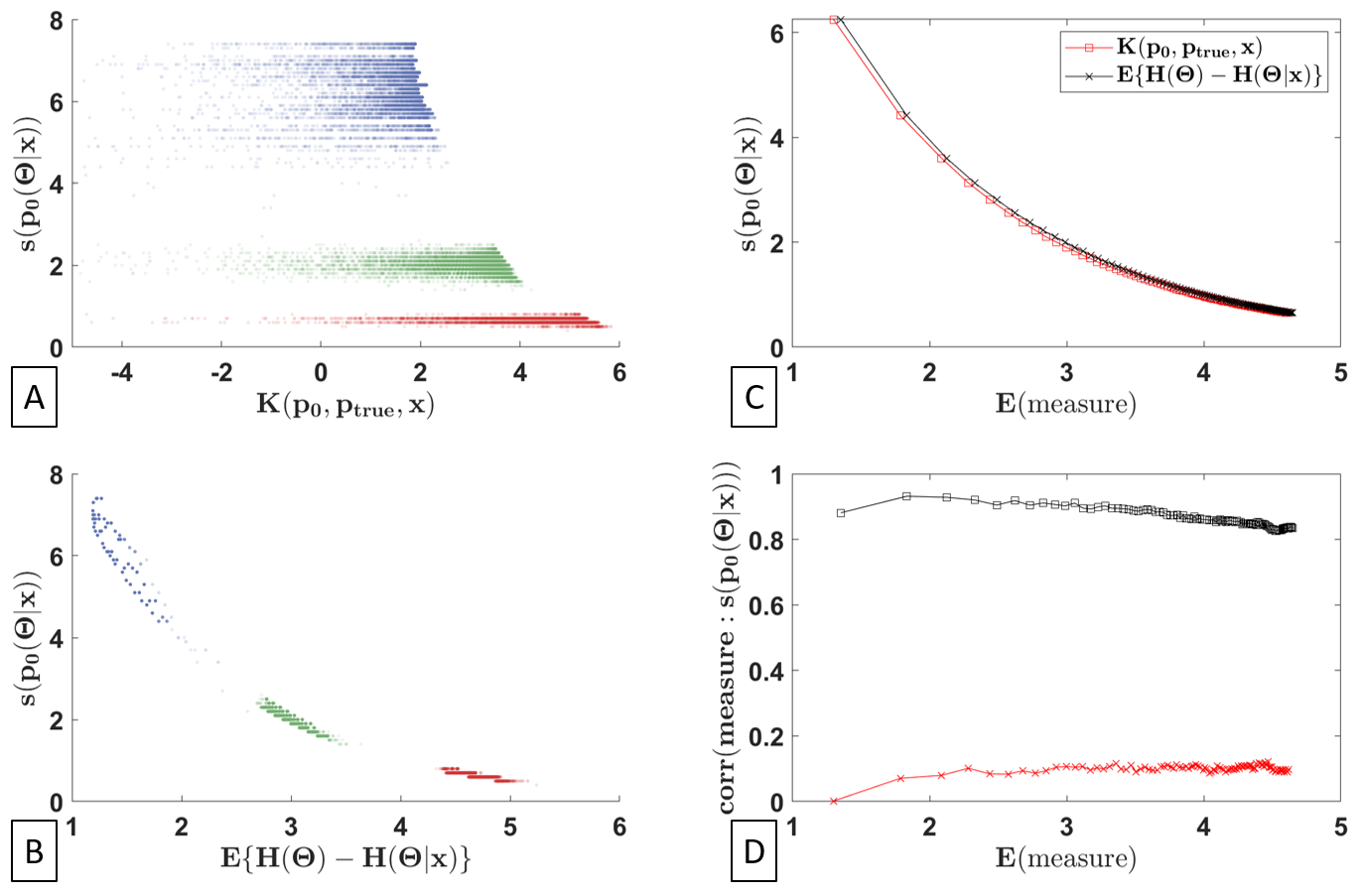}
    }
    \caption{Changes in the values of the two learning measures versus the uncertainty of the inspector's belief in $\Theta$. The uncertainty was estimated by the smallest span around $\tilde{\Theta}$ in $\vartheta$ which included 30\% of the integral of the belief distribution. [A] and [B] show the result for each of the 10,000 cases, after 1, 10, and 100 measurements in blue, green, red respectively. [C] depicts the mean MAP uncertainty versus both learning measures. [D] gives the linear correlation between the belief uncertainty and the two measures after each measurement.}
    \label{fig:numExample_3}
\end{figure}

\noindent Figures \ref{fig:numExample_1}, \ref{fig:numExample_2}, and \ref{fig:numExample_3} compare the privacy / learning assessments using the method we propose, and the Shannon entropy approach. 10,000 warheads were simulated, with the inspector attempting to infer $\Theta$ for each based on up to 100 measurements. The grey lines in Figures \ref{fig:numExample_1}.A and \ref{fig:numExample_1}.B show the progression of the two learning measures for each individual case with each measurement. The Shannon entropy measure increases with almost every measurement, while our proposed measure increases and decreases as more data is collected but drifts upwards on average. The prior distribution was uniformly distributed so any new data, accurate or otherwise, produces a more peaked posterior distribution and reduces the entropy. Our measure is able to better capture when data produces a less accurate estimate of $\Theta$.

While the average change in both measures correspond almost identically with the average change in the error of $\tilde{\Theta}$, they have very different case-by-case correlation, as shown in Figure \ref{fig:numExample_2}. Our proposed measure shows an almost one-to-one relationship with $\tilde{\Theta}$ for each posterior, as shown in Figures \ref{fig:numExample_2}.A and \ref{fig:numExample_2}.D, and there is a linear correlation of $0.88$ across posteriors. While the entropy measure also increases with each measurement, there is only a linear correlation of $0.45$ across the posteriors. 

A similar relationship exists between the two measures and the inspector's belief uncertainty, but the correlation is reversed, as shown in Figure \ref{fig:numExample_3}. We estimate belief uncertainty as the smallest span around $\tilde{\Theta}$ in $\vartheta$ which included 30\% of the integral of the belief distribution. More work is required to understand these relationships in detail and how the operate for any arbitrary distribution.

\section{Inspector with incorrect likelihood function}\label{sec:incorrectlikelihood}

So far we have assumed that the warhead owner and inspector both understand the protocol and measurement processes perfectly. This was reflected in them both knowing the likelihood function of the data given the warhead properties (Equation~\ref{eq:privacy_likelihood}). We made extensive use of this identity when deriving our privacy measure. 

If we relax this assumption by removing Equation~\ref{eq:privacy_likelihood}, we can examine situations in which the inspector attempts to infer $\Theta$ with an incorrect or incomplete likelihood function. This could reflect a range of warhead-protection strategies, for example, where the measured data is scrambled using a secret key, so that the inspector only has access to the naive, key-less likelihood function.

{\footnotesize
\begin{align}
\mathcal{K}(p_0(\Theta),p_{\text{true}}(\Theta),X) &= \sum_{x\in\mathcal{X}}p_{\text{true}}(x|\theta)\sum_{\theta\in\vartheta}p_{\text{true}}(\theta) \log_2\left(\frac{p_0(\theta|x)}{p_0(\theta)}\right)\\
&= \sum_{x\in\mathcal{X}}\sum_{\theta\in\vartheta}p_{\text{true}}(x,\theta) \log_2\left(\frac{p_0(\theta|x)}{p_0(\theta)}\right)
\end{align}
}

\noindent Using the result in Equation \ref{eq:learning_concise}, we can see the difference between this measure and our original result.
{\footnotesize
\begin{align}
\begin{split}
    &\mathcal{K}(p_0(\Theta),p_{\text{true}}(\Theta),x) \\
    &= \sum_{x\in\mathcal{X}}\sum_{\theta\in\vartheta}\left( p_{\text{true}}(x,\theta) \log_2\left(\frac{p_0(\theta|x)}{p_0(\theta)}\right) - p_{\text{true}}(x,\theta)\log_2\left(\frac{p_{\text{true}}(x|\theta)}{p_0(x)}\right)\right )\\
    &\qquad\qquad\qquad + \text{I}(p_{\text{true}}(\Theta);p_{\text{true}}(X)) + D_{KL}(p_{\text{true}}(X)||p_0(X))
\end{split} \\[0.75em]
\begin{split}
    &= \sum_{x\in\mathcal{X}}\sum_{\theta\in\vartheta}\left( 
    p_{\text{true}}(x,\theta) \log_2\left(\frac{p_0(x|\theta)p_0(x)}{p_{\text{true}}(x|\theta)p_0(x)}\right)\right)\\
    &\qquad\qquad\qquad+ \text{I}(p_{\text{true}}(\Theta);p_{\text{true}}(X)) + D_{KL}(p_{\text{true}}(X)||p_0(X))
\end{split} \\[0.75em]
\begin{split}
    &=\text{I}(p_{\text{true}}(\Theta);p_{\text{true}}(X)) + D_{KL}(p_{\text{true}}(X)||p_0(X)) \\
    &\qquad\qquad\qquad\qquad\qquad\qquad - \E_{\theta\in\vartheta}\left\{D_{KL}(p_{\text{true}}(X|\theta)||p_0(X|\theta))\right\}
\end{split}
\end{align}
}

\noindent An incorrect understanding of the likelihood function can decrease the accuracy of the inspector's beliefs. While the third term is a sum over both $X$ and $\Theta$, it still has the properties of a KL divergence.

To demonstrate this more general case, we return to the example in \ref{sec:num_example}, but the output $X$ is modified using a random variable $Y$, with alphabet $y\in\mathcal{Y}$.  A value for $Y$ is assigned to each measurement of the warheads, as described in Equations \ref{eq:key_formula_1} to \ref{eq:key_formula_3}. The inspector knows the distribution of $Y$ but not its value, so attempts to infer $\Theta$ using the likelihood function in Equation \ref{eq:key_formula_4}  

{\footnotesize
\begin{equation}
    p_{\text{true}}(\Theta=\theta) = \mathcal{N}[\theta,\mu=M,\sigma=S^2 ] \label{eq:key_formula_1}
\end{equation}
\begin{equation}
    p_{\text{true}}(Y=y) = \mathcal{N}[\theta,\mu=20,\sigma=15 ]
\end{equation}
\begin{equation}
    p_{\text{true}}(X=x|\theta,y) = \mathcal{P}oisson[x,\lambda=\theta+y ]\label{eq:key_formula_3}
\end{equation}
\begin{equation}
    p_{\text{true}}(X=x|\theta) = \sum_{y\in\mathcal{Y}} p_{\text{true}}(y)\mathcal{P}oisson[x,\lambda=\theta+y ]\label{eq:key_formula_4}
\end{equation}
}


\begin{figure}
    \centering
    \noindent\makebox[0.45\textwidth]{
    \includegraphics[width=0.45\textwidth,angle=-0,origin=c]{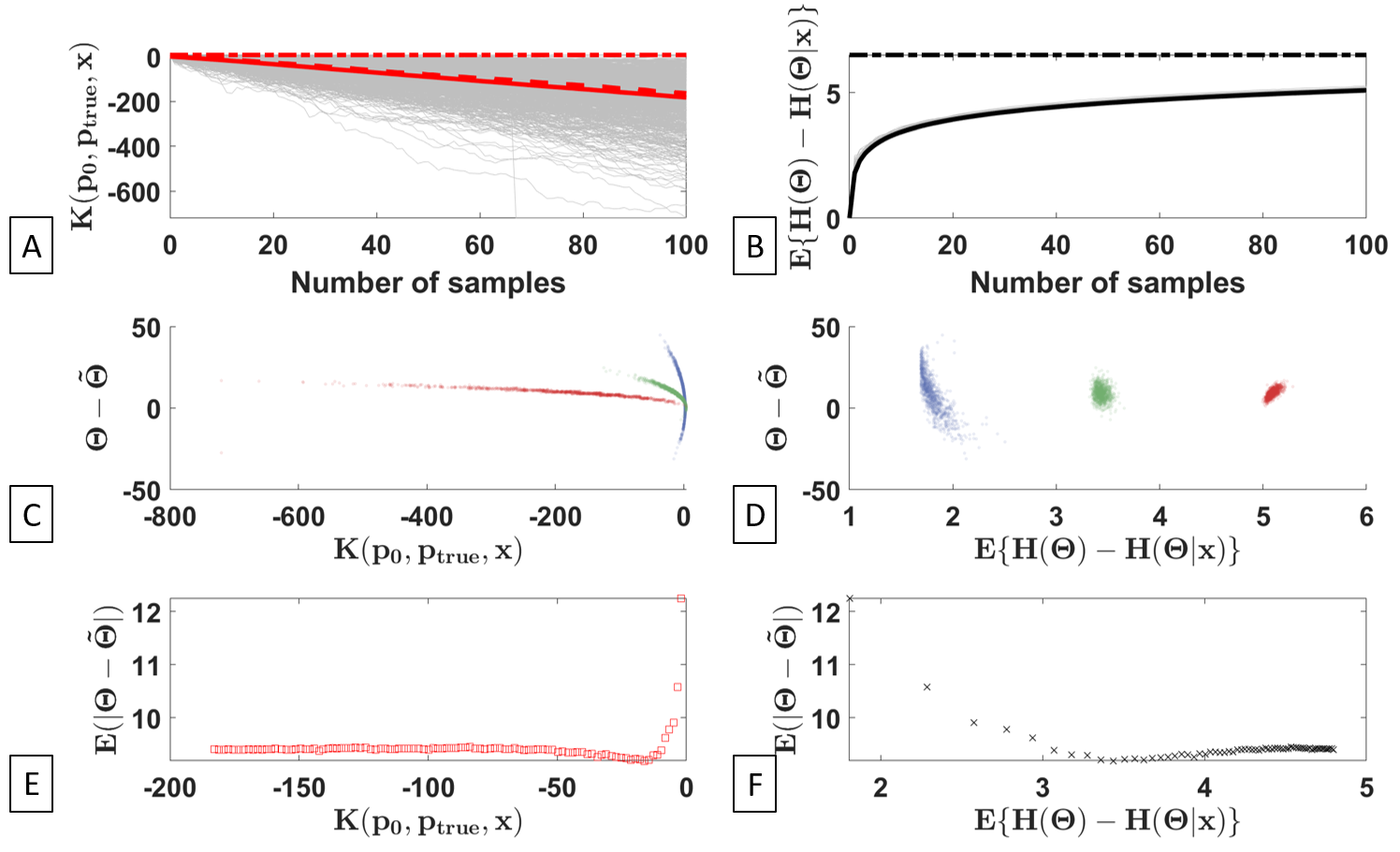}
    }
    \caption{Comparison of our proposed measure and the entropy-based approach when the inspector attempts to infer $\Theta$ using the incorrect likelihood function in Equation~\ref{eq:key_formula_4}. [A] and [B] show the progression of the two measures versus the number of measurements taken over 10,000 simulated cases. The learning measures for each individual case are shown in grey, the mean result by the solid curve, and the median result by the dashed curve. [C] and [D] compare the error in the inspector's MAP estimate of $\Theta$ with the value of the learning measures in each case. [E] and [F] show the mean result of the same.}
    \label{fig:key_example_1}
\end{figure}

This example demonstrates the differences between our proposed measure and the entropy-based approach most clearly. Figure~\ref{fig:key_example_1} compares the two measures with the error in the inspector's MAP estimate of $\Theta$ over 100 measurements of 10,000 warheads. The addition of the secret key causes the inspector's estimate to converge to an incorrect result, on average 9.5 away from the mean. While, the entropy-based measure shows almost the same progression as before, gradually increasing as the inspectors posterior becomes more peaked, our proposed measure decreases in almost all cases, as the inspector's posterior distribution becomes less similar to the correct answer. Because we are most interested in the accuracy, rather than uncertainty, of the inspector's knowledge of the warhead, it is far more useful for a measure of inspector learning / protocol privacy to be sensitive to this change.

\section{Inferring hyperparameters}\label{sec:hyperparameters}

The examples so far have considered the case where an inspector was interested in learning the value of the parameter $\Theta$ for a set of warheads. The inspector may also be interested in the distribution of $\Theta$ itself, governed by the hyperparameters $M$ and $S$. We can assess the privacy of these hyperparameters using the same measure as before.

{\footnotesize
\begin{equation}
\begin{split}
    \mathcal{K}( & p_0(M,S),p_{\text{true}}(M,S),X) \\
    &= \text{I}(p_{\text{true}}(M,S);p_{\text{true}}(X)) + D_{KL}(p_{\text{true}}(X)||p_0(X))
\end{split}
\end{equation}
}

\noindent In the example above, $M$ and $S$ take only one value, as shown in the equation below.

{\footnotesize
\begin{equation}
    p_{\text{true}}(M,S) = \begin{cases}
                                                                1, &\text{if } (M,S) = (M^*\text{, }S^*) \\
                                                                0, &\text{otherwise }
                                                               \end{cases}
\end{equation}
}

\noindent This makes the mutual info / natural learning zero, as there is no uncertainty in $M$ and $S$ if you know $p_{\text{true}}(M,S)$, which is an assumption underlying the mutual information calculation.

{\footnotesize
\begin{equation}
\mathcal{K}(p_0(M,S),p_{\text{true}}(M,S),X) = 0 + D_{KL}(p_{\text{true}}(X|M^*,S^*)||p_0(X))
\end{equation}
}

\section{Summary}

In this paper, we have proposed a new measure for evaluating the privacy of warhead authentication protocols. It allows for equitable comparison of any protocol, which is an improvement over the bespoke measures previously proposed for individual protocols. We have demonstrated that our measure correlates better with the accuracy of a MAP estimate of warhead properties than alternative measures. It can be used to predict the average performance of a protocol beforehand, or a post-fact analysis of the knowledge gained given specific measured data. The measure accommodates different assumptions about which warhead properties need to be considered, and the definition of a measurement can be expanded to include multiple separate measurements, or measurements of multiple objects. This makes the measure flexible and able to be used to compare protocols under a variety of authentication requirements; as long as the same assumptions are used for all of the protocols. These assumptions are a political-technical judgment which must be made before serious comparison of protocols can happen.

While we have focused on privacy and information leakage in a warhead disarmament context, our measure could be used to assess other protocols which deal with private information. It can be applied to situations where a party releases data, knowing the true distribution of the private information, and wishes to predict how much an adversary will improve given inaccurate prior beliefs. For example, if an analysis of a person's medical records were to be released as part of a trial, the measure could be used to calculate how much an adversary would learn about that person's medical history. The adversary's initial belief distribution is based on the properties of the general population (an erroneous $p_0(\Theta)$), while the data is only drawn from the patient ($p_{\text{true}}(\Theta)$). Our measure allows an individual to assess their privacy before consenting to the analysis.

In the introduction to the paper, we discussed the three requirements of authentication protocols: completeness, soundness, and privacy. There is a tension between protocol soundness and privacy, because the former prefers more transparent information to be released to help identify hoaxes, while the latter requires less transparency. Our ongoing work is to develop an equivalent measure for protocol soundness, and understand how it trades-off with privacy.

\bibliography{2018_Warhead_Auth_Privacy.bib}
\bibliographystyle{unsrt} 

\clearpage
\newpage

\end{document}